\documentclass[aps,prl,superscriptaddress,amssymb,twocolumn,showpacs,a4]{revtex4}

\usepackage{epsfig}
\usepackage{graphics}
\usepackage{graphicx}

\begin{document}

\title{Diffusion of liquid domains in lipid bilayer membranes}

\author{Pietro Cicuta}
\affiliation{Cavendish Laboratory and Nanoscience Center, University
of Cambridge,
Cambridge CB3 0HE, U.K.}
\author{Sarah L. Keller}
\affiliation{Departments of Chemistry and Physics, University of
Washington, Seattle, WA 98195-1700, USA}
\author{Sarah L. Veatch}
 \email[]{veatch@cmdr.ubc.ca}
\affiliation{Department of Microbiology and Immunology, University
of British Columbia, Vancouver, BC, V6T1Z4, Canada }

\begin{abstract}
We report diffusion coefficients of micron-scale liquid domains in
giant unilamellar vesicles of phospholipids and cholesterol. The
trajectory of each domain is tracked, and the mean square
displacement grows linearly in time as expected for Brownian motion.
We study domain diffusion as a function of composition and
temperature, and measure how diffusion depends on domain size. We
find mechanisms of domain diffusion which are consistent with
membrane-dominated drag in viscous L$_o$ phases [P.G. Saffman and M.
Delbr\"{u}ck, PNAS 72, 3111 (1975)], and bulk-dominated drag for
less viscous L$_\alpha$ phases [B.D.Hughes et al., J. Fluid Mech.
110, 349 (1981)]. Where applicable, we obtain the membrane viscosity
and report activation energies of diffusion.
\end{abstract}


\maketitle

Diffusion of domains within cell membranes is a highly relevant
biophysical problem. The presence of lipid domains, including rafts,
can affect both short-range (intra-domain) and long-range
(inter-domain) diffusion of membrane
components~\citep{nicolau06,meder06}. Diffusion has been
 observed in live cell membranes~\citep{kusumi05,cell_diff},
although  interpreting results from these complex  systems can be
challenging.

Even in simple model systems, deciphering the diffusion of membrane
inclusions is a long-standing and difficult hydrodynamic
problem~\citep{saffman75,hughes81,clegg85}. Objects that diffuse in
the membrane plane range from small peptides and individual lipids
to large inclusions like protein aggregates and lipid domains. The
first challenge is that, unlike in three dimensional (3D) diffusion,
the size of the diffusing object is not the only length-scale that
enters into the problem. For example, the membrane has finite
thickness, a finite surface area, and often a nonzero curvature.
Secondly, the membrane is composed of macromolecules, which limits
continuum approaches to large objects. Lastly,  the membrane and its
surroundings have different viscosities.  In complex biological
membranes, additional length-scales may be important, such as the
distance between membrane proteins~\citep{saxton94} or the size of
corrals created by the actin cytoskeleton~\citep{kusumi05}.

\begin{figure}[b]
           \epsfig{file=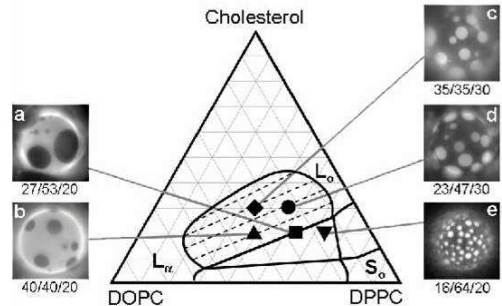,width=7cm}
 \caption{Fluorescence microscopy phase diagram of DOPC/DPPC/cholesterol and vesicle images at
 20$^\circ$C. Semi-quantitative dashed tie-lines cross the L$_\alpha$-L$_o$ coexistence
region~\citep{tielinenote}. Some vesicles studied have a continuous
L$_\alpha$ (bright) phase (a-b) whereas others have a continuous
L$_o$ (dark) phase (c-d). One composition (e) has a continuous dark
L$_o$ phase which may contain both L$_o$ and gel (S$_o$) phase
lipids~\citep{phase_dia}. Vesicle compositions are shown as mol\%
DOPC/DPPC/Chol.
  \label{fig1}}
\end{figure}

In this Letter, we directly measure diffusion of liquid domains in
giant unilamellar vesicles (GUVs) of radius $\simeq20\mu$m as in
Figure~\ref{fig1}. These domains  are micron-scale, circular, span
the lipid bilayer, and undergo Brownian motion.  By measuring
diffusion of bilayer domains over a wide parameter range of more
than one decade in domain radii and three decades in 2D membrane
viscosities, we probe the two limiting models of
Saffman-Delbr\"{u}ck~\cite{saffman75} and Hughes et
al.~\cite{hughes81}.  We find a cross-over between the two models
which would not have been predicted from previous monolayer
results~\cite{mcconnell93}. In the cases where our data are well fit
by the Saffman-Delbr\"{u}ck equation, we are able to extract
viscosities of lipid phases and diffusion activation energies.

Domains move in a background phase with two dimensional (2D)
membrane viscosity ($\eta''$). The diffusion coefficient of a
membrane inclusion was originally described by Saffman and
Delbr\"{u}ck~\cite{saffman75}:
\begin{eqnarray}
   D(r)\,=\,\frac{k_B T}{4 \pi \eta''}
   \left[ \log\left(\frac{\eta''}{\eta_w}\frac{1}{r}\right) - \gamma +\frac{1}{2}
   \right],
 \label{eq1}
\end{eqnarray}
where  $r$ is the radius of the inclusion, $\gamma=0.5772$ and we
have chosen boundary conditions appropriate for liquid domains in a
liquid membrane to yield the factor of $1/2$. A key parameter in the
hydrodynamics of this system is the lengthscale $\lambda_0$ defined
by the ratio of the membrane $\eta''$ to the 3D bulk viscosity of
water ($\eta_w$) such that $\lambda_0=\eta''/\eta_w$. Eq.~\ref{eq1}
is expected to hold for $r<\lambda_0$, i.e. small domains and/or
large membrane viscosity. Later work confirmed this calculation and
derived a result for the opposite limit of
$\lambda_0<r$~\cite{hughes81}:
\begin{eqnarray}
   D(r)\,=\,\frac{k_B T}{16 \eta_w}\frac{1}{r}.
 \label{eq1b}
\end{eqnarray}
It is important to notice that for $\lambda_0<r$ the diffusion
coefficient is more strongly dependent on the inclusion's radius,
but is independent of the membrane viscosity. This case was verified
experimentally through observations of micron-scale domains in
monolayers~\cite{mcconnell93}.
 Other theoretical work has addressed
different inclusion shapes as well as large
domains~\cite{diff_theory}.

\begin{figure}[t]
           \epsfig{file=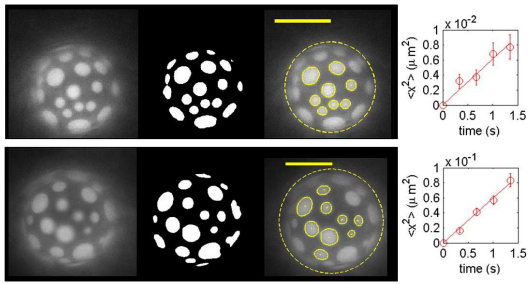,width=7cm}
 \caption{ Greyscale fluorescence
 images (left)
 are filtered and thresholded (middle). White regions are identified as
 domains.
  Those within a specified size range, ellipticity and distance from the edge are retained.
 Circles (right image) identify those domains successfully identified through 5 successive frames.
 Mean square displacement data for domains with radii of 1-1.5$\mu$m are shown at right.
 Both vesicles have composition 1:2 DOPC:DPPC + 30\%Chol, T=10$^\circ$C (top)  and
  20$^\circ$C (bottom).  Note the factor of ten
 difference in diffusion coefficients. The scale bar is 40$\mu$m.
    \label{fig2}}
\end{figure}


Spherical giant unilamellar vesicles (GUVs; 30-100$\mu$m diameter)
are made by electroformation~\citep{Angelova92} of a ternary mixture
of cholesterol with phospholipids of both high (DPPC; di(16:0)PC)
and low (DOPC; di(18:1)PC) melting temperatures. Materials and
methods have been described previously \citep{phase_dia}. The
vesicle membranes are initially uniform at high temperature, and
phase separate into two liquid phases when vesicle suspensions are
placed on a pre-cooled microscope stage. The less viscous L$_\alpha$
phase is labeled by fluorescent dye (Texas Red-DPPE). The
composition and viscosity of the two phases depend on the
composition and temperature of the entire vesicle. With time,
domains coalesce, allowing us to probe a range of domain sizes at
constant temperature.

We probe five lipid compositions in the ternary system of
DOPC/DPPC/Chol, as in Figure~\ref{fig1}. In previous microscopy and
$^2$H NMR measurements, we established that vesicles with these five
compositions separate into a liquid-ordered (L$_o$) phase rich in
the saturated lipid DPPC and a liquid crystalline (L$_\alpha$) phase
rich in the unsaturated lipid DOPC \citep{phase_dia,veatch04}. Two
of the five compositions contain a continuous bright L$_\alpha$
phase (Figure \ref{fig1}a-b), and two contain a continuous dark
L$_o$ phase (Figure \ref{fig1}c-d).  One composition falls within a
three phase region (Figure \ref{fig1}e.)  The presence of three
phases is clear in $^2$H NMR experiments (manuscript in
preparation), but is difficult to detect by microscopy. We probe the
viscosity of the continuous phase by tracking domains of the
minority phase.

Membrane domains are identified by an image processing program
written in Matlab (Figure~\ref{fig2}). A Gaussian filter is applied
to images before thresholding, to identify features in the size
range of domains. Almost no domains are lost by this algorithm.
Domains are accepted if: a) the diameter falls between a minimum (2
pixel) and maximum value; b) the shape is circular, such that all
points in the domain perimeter lie within 20\% of the mean domain
radius; and c) the center of mass lies within a circle defined by
0.8 of the vesicle radius. Since all domains are round (shape
fluctuations are minimal), these criteria discriminate against
occasional problems arising from image analysis filtering (for
example, two domains very close to each other will not be accepted).

A separate program tracks domain trajectories with logic similar to
existing codes, i.e. by matching a domain with the nearest feature
in the next image \cite{tracknote}.  Average diffusion is subtracted
to yield unbiased domain motion. Matching generates no false
positives but does not have a perfect success rate. We therefore
divide each movie (typically 100 frames at 0.34 s/frame) into 20
sets of 5 frames over which most domains are tracked successfully.
Domain size does not change over this period. The average of
vertical and horizontal mean square displacements (MSD) is linear
with time $t$ and fit to $<x^2>=2D(r)t$ as expected for diffusion.
Over five frames, the MSD is $\leq0.03\mu$m$^2$ which is much
smaller than the particle separation, and we see no effect of domain
packing.
\begin{figure}[t]
           \epsfig{file=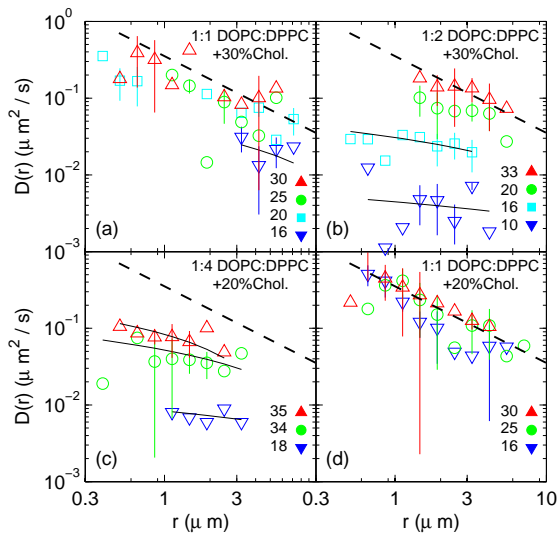,width=7.5cm}
 \caption{Diffusion coefficients vary with domain radius.
 Solid lines show fits to a logarithmic dependence on domain
 size as in the Saffman-Delbr\"{u}ck Eq.~\ref{eq1}. These fits have
 only one free parameter, as discussed in the text.  Symbols
 identify increasing temperatures, recorded each plot.
 Dashed lines are fits to Eq.~\ref{eq2} with no free
 parameters. Error bars report standard deviations for sets with
$\geq$3 measurements. \label{fig3}}
\end{figure}


Figure~\ref{fig3} shows diffusion coefficients as a function of
domain size. The data has been culled to report only sets in which
wide ranges of domain sizes are observed for any fixed temperature
and composition.  The dashed lines in Figure~\ref{fig3} show the
asymptotic $1/r$ behavior given in Eq.~\ref{eq1b}.  Within error,
all data fall on or below this theoretical upper bound in diffusion
coefficient. Eq.~\ref{eq1b} is independent of membrane viscosity and
holds when membrane viscosity is low, or domain radius is large i.e.
when $\lambda_0\ll r$~\cite{hughes81}. For the low viscosity
$L_\alpha$ phase (e.g. Figure~\ref{fig3}d), it can be seen that the
conditions of low membrane viscosity and large domain radius are met
through most of the temperature range, because most data fall along
the dashed line.

All data below the dashed line in Figure~\ref{fig3} correspond to
membranes with high viscosity, notably L$_o$ phases at low
temperatures.  We have chosen to fit our data to the
Saffman-Delbr\"{u}ck equation, which should hold when membrane
viscosity is high.  It is clear that this set of data does not have
a $D(r)\sim r^{-1}$ dependence, and  we find instead reasonable fits
to Eq.~\ref{eq1} with a single fitting parameter ($\eta''$). Since
our domain radii are limited to $\geq$0.5$\mu$m by our optical
resolution and $\leq$10$\mu$m by our vesicle diameters, we cannot
prove that Eq.~\ref{eq1} is the only expression that could fit our
data. Nevertheless, fitting to Eq.~\ref{eq1} allows us to extract
$\eta''$ for high viscosity membranes.

There have been multiple experiments designed to test the
logarithmic form of
Eq.~\ref{eq1}~\citep{gambin06,clegg85,mcconnell93}, and its range of
applicability is still controversial. For example, recent work
asserts that individual proteins diffuse with a stronger size
dependence, as $D(r)\sim r^{-1}$~\cite{gambin06}, due to a
break-down of the continuum approximation of the membrane for small
inclusions~\cite{gambin06,lee03}. Here we explore domain radii well
within the continuum limit ($r\gg r_{\mathrm{single\
molecule}}\sim0.5$nm).

Our results can be extrapolated to estimate the diffusion rate of
raft domains in cell membranes. Lipid rafts are reported to have
diameters of 10-100nm~\citep{raft_size}. This length-scale falls
within the regime where the Saffman-Delbr\"{u}ck equation should
apply. We calculate diffusion coefficients for 10-100nm domains in
our system to be between $3\times10^{-3}$ and
$1.5\times10^{-1}\mu$m$^2$/sec. These values differ from those
extrapolated from single molecule measurements using the
Saffman-Delbr\"{u}ck equation~\cite{nicolau06}.  Given the
suggestion that single molecule measurements do not probe the
continuum limit, it may be more valid to estimate raft diffusion
coefficients by extrapolating down from large domains rather than up
from single molecules.

\begin{figure}[t]
           \epsfig{file=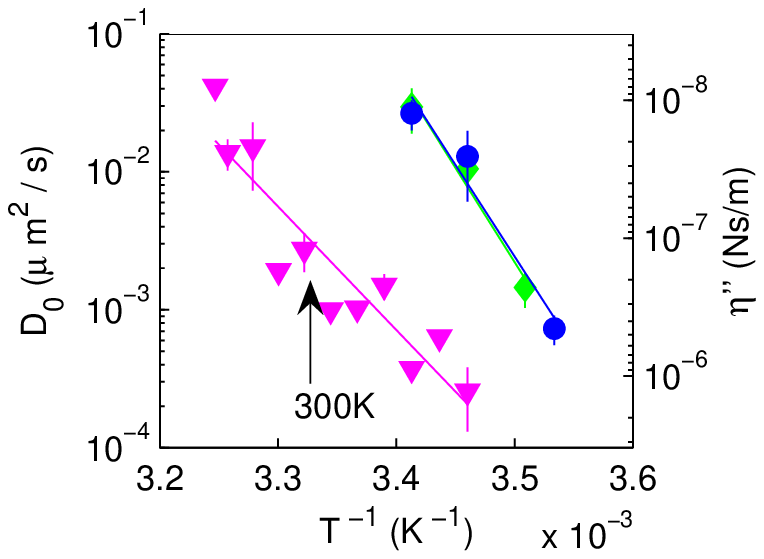,width=5.5cm}  
 {\scriptsize
           \begin{tabular}{ccc}
  Composition & Symbol  & \,\,\,\,\,E$_a $  (kJ/mol)  \\
  \hline
  1:1 DOPC/DPPC + 30\%Chol & $\blacklozenge$  & 248$\pm$2  \\
  1:2 DOPC/DPPC + 30\%Chol&  $\bullet$ & 252$\pm$2  \\
  1:4 DOPC/DPPC + 20\%Chol & $\blacktriangledown$ & 123$\pm$8  \\
  \hline

\end{tabular}
}
 \caption{The diffusion coefficient $D_0$ and the 2D membrane viscosity $\eta''$
 as a function of temperature in membranes with majority L$_o$ and S$_o$-L$_o$ phases.
 Solid lines are
 fit to  $D_0\sim \exp(-E_a/k_BT)$, with $E_a$ values shown in the table.   \label{fig5}}
\end{figure}

Our analysis of the culled data set in Figure~\ref{fig3} shows that
high viscosity membranes produce data that fall well below the
dashed line in Figure~\ref{fig3} and that fit Eq. \ref{eq1}
reasonably well. We conclude that any remaining unplotted data that
fall well below the dashed line should also fit Eq.~\ref{eq1}. We
use this data to yield a size-independent $D_0$ using:
\begin{eqnarray}
   D(r)\,&=&\,D_0
   \left[ \log\left(\frac{\lambda_0}{r}\right) - 0.0772
   \right],\label{eq2a}\\
    \mathrm{with} \,\, D_0\,&=&\,\frac{k_B T}{4 \pi \eta''}.
 \label{eq2}
\end{eqnarray}
Figure~\ref{fig5} shows a plot of the resulting $D_0$ and 2D
membrane viscosity vs. temperature (T).  We find that the L$_o$
phase viscosities for membranes of 1:1 DOPC/DPPC + 30\%Chol and 1:2
DOPC/DPPC + 30\%Chol are similar, suggesting that L$_o$ viscosities
are not highly composition dependent. In contrast, viscosities for
membranes of 1:4 DOPC/DPPC + 20\%Chol are much larger, consistent
with these membranes falling within the three phase region in
Figure~\ref{fig1}.

At 22$^\circ$C, we find 2D membrane viscosities for the $L_o$ phase
of $10^{-8}\leq \eta''\mathrm{(Ns/m)} \leq 5\times10^{-7}$.  Surface
shear rheometry finds monolayer viscosities on the order of
$10^{-8}$ to $10^{-6}$(Ns/m) only in liquid condensed
phases~\cite{fuller99}, which is consistent with tight packing of
lipids in bilayer L$_o$ phases. In contrast, the 2D membrane
viscosity of the L$_\alpha$ phase is low, and results in D$\sim
r^{-1}$.  In monolayers, the same D$\sim r^{-1}$ dependence is found
for solid domains diffusing across a liquid
background~\cite{mcconnell93}.

In the membrane literature, a 3D membrane viscosity, $\eta_{3D}$, is
defined as $\eta_{3D}\simeq \eta'' / h$, where $h$ is the bilayer
thickness. Assuming $h=3.3$nm, we find $3\leq \eta_{3D}\mathrm{(Pa\,
s)} \leq 150$, on the order of ~\cite{gambin06,vaz87} or greater
than~\citep{visc_refs} published values for model membranes.
However, the relation between $\eta_{3D}$ and $\eta''$ is not exact,
because lipids anchored to the interface differ from a thin
homogeneous layer. Indeed, the lipid headgroups often determine the
membrane viscosity~\cite{clegg85}. This is not always appreciated,
and may be a source of ambiguity in discussions of the
Saffman-Delbr\"{u}ck model~\cite{gambin06}.

Figure~\ref{fig5}  demonstrates that domains diffuse in membranes of
high viscosity via an activated process. If the 2D membrane
viscosity ($\eta''$) were independent of T, we would expect
$D_0(T)\sim T$. Instead, we find a better fit for $\log(D_0(T))\sim
-T^{-1}$, consistent with an activation energy $E_a$ for diffusion
such that $D_0\sim \exp(-E_a/k_BT)$.  The data in Figure 4 follow
Arrhenius behavior even though a gel phase emerges at low
temperature for some mixtures.  Composition of the L$_o$ phase
varies only slightly with temperature~\cite{phase_dia}. Activation
energies for individual lipids~\citep{scheidt05,lindblom03,vaz92}
have been attributed to the energy required to hop into an available
free volume~\cite{clegg85,vaz92}. Larger particles such as protein
aggregates yield lower apparent activation energies~\cite{lee03}.
Fig.~\ref{fig5} lists activation energies for domains diffusing in
L$_o$ phases. We find activation energies greater than those
reported for single molecules in similar membranes, including
DPPC/Chol membranes at high temperature
(30-80kJ/mol)~\citep{scheidt05}, as well as L$_o$ lipids in phase
separated DOPC/DPPC/Chol membranes at low temperature
($\sim80$kJ/mol)~\citep{oradd05}.

In summary, we present a simple method for quantifying the movement
of domains in membranes with coexisting liquid phases.  We find that
domains diffuse via Brownian motion, and that diffusion rates are
described by different models under different experimental
conditions. At high temperatures and in membranes with a continuous
L$_\alpha$ phase, membrane viscosity is low, diffusion constants are
independent of membrane properties, and domains diffuse with a
radial dependence of $D\sim1/r$.  In membranes with a higher
viscosity continuous phase, domain movement does depend on membrane
physical properties and the radial dependence can be fit by a
Saffman-Delbr\"{u}ck model with $D\sim$log$(1/r)$. For these
membranes, we determine 2D viscosities and report activation
energies for domain diffusion.

\begin{acknowledgments}
PC was funded by the Oppenheimer Fund, EPSRC, and the
Cavendish-KAIST Cooperative Research Program of MoST Korea. SLK
acknowledges an NSF CAREER award and a Cottrell Scholar award.  SLV
acknowledges a grant from the Cancer Research Institute. We thank
Imran Hasnain for help with image analysis, and Klaus Gawrisch for
helpful conversations.
\end{acknowledgments}

\bibliography{diff_bibdata}

\begin{thebibliography}{25}
\expandafter\ifx\csname natexlab\endcsname\relax\def\natexlab#1{#1}\fi
\expandafter\ifx\csname bibnamefont\endcsname\relax
  \def\bibnamefont#1{#1}\fi
\expandafter\ifx\csname bibfnamefont\endcsname\relax
  \def\bibfnamefont#1{#1}\fi
\expandafter\ifx\csname citenamefont\endcsname\relax
  \def\citenamefont#1{#1}\fi
\expandafter\ifx\csname url\endcsname\relax
  \def\url#1{\texttt{#1}}\fi
\expandafter\ifx\csname urlprefix\endcsname\relax\def\urlprefix{URL }\fi
\providecommand{\bibinfo}[2]{#2}
\providecommand{\eprint}[2][]{\url{#2}}

\bibitem[{\citenamefont{Nicolau et~al.}(2006)\citenamefont{Nicolau, Burrage,
  Parton, and Hancock}}]{nicolau06}
\bibinfo{author}{\bibfnamefont{D.~V.~J.} \bibnamefont{Nicolau}},
  \bibinfo{author}{\bibfnamefont{K.}~\bibnamefont{Burrage}},
  \bibinfo{author}{\bibfnamefont{R.~G.} \bibnamefont{Parton}},
  \bibnamefont{and} \bibinfo{author}{\bibfnamefont{J.~F.}
  \bibnamefont{Hancock}}, \bibinfo{journal}{Mol. Cell Biol.}
  \textbf{\bibinfo{volume}{26}}, \bibinfo{pages}{313} (\bibinfo{year}{2006}).

\bibitem[{\citenamefont{Meder et~al.}(2006)\citenamefont{Meder, Moreno,
  Verkade, Vaz, and Simons}}]{meder06}
\bibinfo{author}{\bibfnamefont{D.}~\bibnamefont{Meder}},
  \bibinfo{author}{\bibfnamefont{M.~J.} \bibnamefont{Moreno}},
  \bibinfo{author}{\bibfnamefont{P.}~\bibnamefont{Verkade}},
  \bibinfo{author}{\bibfnamefont{W.~L.} \bibnamefont{Vaz}}, \bibnamefont{and}
  \bibinfo{author}{\bibfnamefont{K.}~\bibnamefont{Simons}},
  \bibinfo{journal}{Proc. Natl. Acad. Sci. USA.}
  \textbf{\bibinfo{volume}{103}}, \bibinfo{pages}{329} (\bibinfo{year}{2006}).

\bibitem[{\citenamefont{Kusumi and Suzuki}(2005)}]{kusumi05}
\bibinfo{author}{\bibfnamefont{A.}~\bibnamefont{Kusumi}} \bibnamefont{and}
  \bibinfo{author}{\bibfnamefont{K.}~\bibnamefont{Suzuki}},
  \bibinfo{journal}{Biochim. Biophys. Acta.} \textbf{\bibinfo{volume}{1746}},
  \bibinfo{pages}{234} (\bibinfo{year}{2005}).

\bibitem[{cel()}]{cell_diff}
\bibinfo{note}{A. K. Kenworthy, B. J. Nichols, C. L. Remmert, G. M. Hendrix, M.
  Kumar, J. Zimmerberg, and J. Lippincott-Schwartz, J. Cell Biol. 165, 735
  (2004); Y. Chen, B. Yang, and K. Jacobson, Lipids 39, 1115 (2004); K. Bacia,
  D. Scherfeld, N. Kahya, and P. Schwille, Biophys. J. 87, 1034 (2004).}

\bibitem[{\citenamefont{Saffman and Delbr\"{u}ck}(1975)}]{saffman75}
\bibinfo{author}{\bibfnamefont{P.}~\bibnamefont{Saffman}} \bibnamefont{and}
  \bibinfo{author}{\bibfnamefont{M.}~\bibnamefont{Delbr\"{u}ck}},
  \bibinfo{journal}{Proc. Natl. Acad. Sci.} \textbf{\bibinfo{volume}{72}},
  \bibinfo{pages}{3111} (\bibinfo{year}{1975}).

\bibitem[{\citenamefont{B.D.Hughes et~al.}(1981)\citenamefont{B.D.Hughes,
  B.A.Pailthorpe, and L.R.White}}]{hughes81}
\bibinfo{author}{\bibnamefont{B.D.Hughes}},
  \bibinfo{author}{\bibnamefont{B.A.Pailthorpe}}, \bibnamefont{and}
  \bibinfo{author}{\bibnamefont{L.R.White}}, \bibinfo{journal}{J. Fluid Mech.}
  \textbf{\bibinfo{volume}{110}}, \bibinfo{pages}{349} (\bibinfo{year}{1981}).

\bibitem[{\citenamefont{Clegg and Vaz}(1982)}]{clegg85}
\bibinfo{author}{\bibfnamefont{R.~M.} \bibnamefont{Clegg}} \bibnamefont{and}
  \bibinfo{author}{\bibfnamefont{W.~L.~C.} \bibnamefont{Vaz}}, in
  \emph{\bibinfo{booktitle}{Progress in Protein-Lipid Interactions}}, edited by
  \bibinfo{editor}{\bibnamefont{Watts}} \bibnamefont{and}
  \bibinfo{editor}{\bibnamefont{{De Pont}}} (\bibinfo{publisher}{Elsevier},
  \bibinfo{year}{1982}).

\bibitem[{\citenamefont{Saxton}(1994)}]{saxton94}
\bibinfo{author}{\bibfnamefont{M.~J.} \bibnamefont{Saxton}},
  \bibinfo{journal}{Biophys. J.} \textbf{\bibinfo{volume}{66}},
  \bibinfo{pages}{394} (\bibinfo{year}{1994}).

\bibitem[{tie()}]{tielinenote}
\bibinfo{note}{Quantitative tie-lines are determined by $^2$H NMR (manuscript
  in preparation). Fluorescence microscopy phase boundaries differ from those
  determined by $^2$H NMR due to the presence of probe lipid.}

\bibitem[{pha()}]{phase_dia}
\bibinfo{note}{S.L. Veatch and S.L. Keller, Phys. Rev. Lett. 89, 268101 (2002);
  S. L. Veatch and S. L. Keller, Biophys. J. 85, 3074 (2003); S. L. Veatch and
  S. L. Keller, Biochim. Biophys. Acta. 1746, 172 (2005).}

\bibitem[{\citenamefont{Klingler and McConnell}(1993)}]{mcconnell93}
\bibinfo{author}{\bibfnamefont{J.~F.} \bibnamefont{Klingler}} \bibnamefont{and}
  \bibinfo{author}{\bibfnamefont{H.~M.} \bibnamefont{McConnell}},
  \bibinfo{journal}{J. Phys. Chem.} \textbf{\bibinfo{volume}{97}},
  \bibinfo{pages}{6096} (\bibinfo{year}{1993}).

\bibitem[{dif()}]{diff_theory}
\bibinfo{note}{H.A. Stone and A. Ajdari, J. Fluid Mech. 369, 151 (1998); A.J.
  Levine, T.B. Liverpool, and F.C. MacKintosh, Phys. Rev. Lett. 93, 038102
  (2004); \emph{ibid}, Phys. Rev. E 69, 021503 (2004).}

\bibitem[{\citenamefont{Angelova et~al.}(1992)\citenamefont{Angelova, Soleau,
  Meleard, Faucon, and Bothorel}}]{Angelova92}
\bibinfo{author}{\bibfnamefont{M.~I.} \bibnamefont{Angelova}},
  \bibinfo{author}{\bibfnamefont{S.}~\bibnamefont{Soleau}},
  \bibinfo{author}{\bibfnamefont{P.}~\bibnamefont{Meleard}},
  \bibinfo{author}{\bibfnamefont{J.~F.} \bibnamefont{Faucon}},
  \bibnamefont{and} \bibinfo{author}{\bibfnamefont{P.}~\bibnamefont{Bothorel}},
  \bibinfo{journal}{Progr. Colloid Polym. Sci.} \textbf{\bibinfo{volume}{89}},
  \bibinfo{pages}{127} (\bibinfo{year}{1992}).

\bibitem[{\citenamefont{Veatch et~al.}(2004)\citenamefont{Veatch, Polozov,
  Gawrisch, and Keller}}]{veatch04}
\bibinfo{author}{\bibfnamefont{S.~L.} \bibnamefont{Veatch}},
  \bibinfo{author}{\bibfnamefont{I.~V.} \bibnamefont{Polozov}},
  \bibinfo{author}{\bibfnamefont{K.}~\bibnamefont{Gawrisch}}, \bibnamefont{and}
  \bibinfo{author}{\bibfnamefont{S.~L.} \bibnamefont{Keller}},
  \bibinfo{journal}{Biophys. J.} \textbf{\bibinfo{volume}{86}},
  \bibinfo{pages}{2910} (\bibinfo{year}{2004}).

\bibitem[{tra()}]{tracknote}
\bibinfo{note}{J. C. Crocker and D. G. Grier, J. Coll. Int. Sci. 179, 298
  (1996); We track only features identified on the first frame of a set of
  images. We do not need to track ``new'' features because movie segments are
  short and features in 2D remain in the field of view. In contrast, colloidal
  particles in 3D can leave the field of focus.}

\bibitem[{\citenamefont{Gambin et~al.}(2006)\citenamefont{Gambin,
  {Lopez-Esparza}, Reffay, Sierecki, Gov, Genest, Hodges, and
  Urbach}}]{gambin06}
\bibinfo{author}{\bibfnamefont{Y.}~\bibnamefont{Gambin}},
  \bibinfo{author}{\bibfnamefont{R.}~\bibnamefont{{Lopez-Esparza}}},
  \bibinfo{author}{\bibfnamefont{M.}~\bibnamefont{Reffay}},
  \bibinfo{author}{\bibfnamefont{E.}~\bibnamefont{Sierecki}},
  \bibinfo{author}{\bibfnamefont{N.~S.} \bibnamefont{Gov}},
  \bibinfo{author}{\bibfnamefont{M.}~\bibnamefont{Genest}},
  \bibinfo{author}{\bibfnamefont{R.~S.} \bibnamefont{Hodges}},
  \bibnamefont{and} \bibinfo{author}{\bibfnamefont{W.}~\bibnamefont{Urbach}},
  \bibinfo{journal}{Proc. Natl. Acad. Sci.} \textbf{\bibinfo{volume}{103}},
  \bibinfo{pages}{2098} (\bibinfo{year}{2006}).

\bibitem[{\citenamefont{Lee et~al.}(2003)\citenamefont{Lee, Revington, Dunn,
  and Petersen}}]{lee03}
\bibinfo{author}{\bibfnamefont{C.}~\bibnamefont{Lee}},
  \bibinfo{author}{\bibfnamefont{M.}~\bibnamefont{Revington}},
  \bibinfo{author}{\bibfnamefont{S.}~\bibnamefont{Dunn}}, \bibnamefont{and}
  \bibinfo{author}{\bibfnamefont{N.}~\bibnamefont{Petersen}},
  \bibinfo{journal}{Biophys. J.} \textbf{\bibinfo{volume}{84}},
  \bibinfo{pages}{1756} (\bibinfo{year}{2003}).

\bibitem[{raf()}]{raft_size}
\bibinfo{note}{S. J. Plowman, C. Muncke, R. G. Parton, and J. F. Han- cock,
  Proc. Nat. Acad. Sci. 102, 15500 (2005). A. Pralle, P. Keller, E. L. Florin,
  K. Simons, and J. K. Horber, J. Cell Biol. 148, 997 (2000). P. Sharma, R.
  Varma, R. C. Sarasij, Ira, K. Gousset, G. Krishnamoorthy, M. Rao, and S.
  Mayor, Cell 116, 577 (2004).}

\bibitem[{\citenamefont{Brooks et~al.}(1999)\citenamefont{Brooks, Fuller,
  Curtis, and Robertson}}]{fuller99}
\bibinfo{author}{\bibfnamefont{C.~F.} \bibnamefont{Brooks}},
  \bibinfo{author}{\bibfnamefont{G.~G.} \bibnamefont{Fuller}},
  \bibinfo{author}{\bibfnamefont{C.~W.} \bibnamefont{Curtis}},
  \bibnamefont{and} \bibinfo{author}{\bibfnamefont{C.~R.}
  \bibnamefont{Robertson}}, \bibinfo{journal}{Langmuir}
  \textbf{\bibinfo{volume}{15}}, \bibinfo{pages}{2450} (\bibinfo{year}{1999}).

\bibitem[{\citenamefont{Vaz et~al.}(1987)\citenamefont{Vaz, Stümpel, Hallman,
  Gambacorta, and Rosa}}]{vaz87}
\bibinfo{author}{\bibfnamefont{W.}~\bibnamefont{Vaz}},
  \bibinfo{author}{\bibfnamefont{J.}~\bibnamefont{Stümpel}},
  \bibinfo{author}{\bibfnamefont{D.}~\bibnamefont{Hallman}},
  \bibinfo{author}{\bibfnamefont{A.}~\bibnamefont{Gambacorta}},
  \bibnamefont{and} \bibinfo{author}{\bibfnamefont{M.~D.} \bibnamefont{Rosa}},
  \bibinfo{journal}{Eur. Biophys. J.} \textbf{\bibinfo{volume}{15}},
  \bibinfo{pages}{111} (\bibinfo{year}{1987}).

\bibitem[{vis()}]{visc_refs}
\bibinfo{note}{R. Peters and R. Cherry, Proc. Natl. Acad. Sci. USA. 14, 4317
  (1982); C. Chang, H. Takeuchi, T. Ito, K. Machida, and S. Ohnishi, J.
  Biochem. 90, 997 (1981).}

\bibitem[{\citenamefont{Scheidt et~al.}(2005)\citenamefont{Scheidt, Huster, and
  Gawrisch}}]{scheidt05}
\bibinfo{author}{\bibfnamefont{H.~A.} \bibnamefont{Scheidt}},
  \bibinfo{author}{\bibfnamefont{D.}~\bibnamefont{Huster}}, \bibnamefont{and}
  \bibinfo{author}{\bibfnamefont{K.}~\bibnamefont{Gawrisch}},
  \bibinfo{journal}{Biophys. J.} \textbf{\bibinfo{volume}{89}},
  \bibinfo{pages}{2504} (\bibinfo{year}{2005}).

\bibitem[{\citenamefont{Filippov et~al.}(2003)\citenamefont{Filippov, Orädd,
  and Lindblom}}]{lindblom03}
\bibinfo{author}{\bibfnamefont{A.}~\bibnamefont{Filippov}},
  \bibinfo{author}{\bibfnamefont{G.}~\bibnamefont{Orädd}}, \bibnamefont{and}
  \bibinfo{author}{\bibfnamefont{G.}~\bibnamefont{Lindblom}},
  \bibinfo{journal}{Biophys. J.} \textbf{\bibinfo{volume}{84}},
  \bibinfo{pages}{3079} (\bibinfo{year}{2003}).

\bibitem[{\citenamefont{Almeida et~al.}(1992)\citenamefont{Almeida, Vaz, and
  Thompson}}]{vaz92}
\bibinfo{author}{\bibfnamefont{P.~F.~F.} \bibnamefont{Almeida}},
  \bibinfo{author}{\bibfnamefont{W.~L.~C.} \bibnamefont{Vaz}},
  \bibnamefont{and} \bibinfo{author}{\bibfnamefont{T.~E.}
  \bibnamefont{Thompson}}, \bibinfo{journal}{Biochemistry}
  \textbf{\bibinfo{volume}{31}}, \bibinfo{pages}{6739} (\bibinfo{year}{1992}).

\bibitem[{\citenamefont{Oradd et~al.}(2005)\citenamefont{Oradd, Westerman, and
  G.}}]{oradd05}
\bibinfo{author}{\bibfnamefont{G.}~\bibnamefont{Oradd}},
  \bibinfo{author}{\bibfnamefont{P.~W.} \bibnamefont{Westerman}},
  \bibnamefont{and} \bibinfo{author}{\bibfnamefont{L.}~\bibnamefont{G.}},
  \bibinfo{journal}{Biophys. J.} \textbf{\bibinfo{volume}{89}},
  \bibinfo{pages}{315} (\bibinfo{year}{2005}).

\end{thebibliography}

\end{document}